\documentclass{article}
\usepackage{epsfig}

\title{\large Mott-Hagedorn Resonance Gas and Lattice QCD Results\thanks{Talk presented at the Strangeness in Quark Matter SQM2011 Conference, Cracow, Poland, September 18-24 2011}}%
\author{\normalsize L.~Turko$^1$\footnote{\emph{e-mail address}: turko@ift.uni.wroc.pl},
D. Blaschke$^{1,2}$\footnote{\emph{e-mail address}: blaschke@ift.uni.wroc.pl},
D.~Prorok$^1$\footnote{\emph{e-mail address}: prorok@ift.uni.wroc.pl} and
J.~Berdermann$^3$\footnote{\emph{e-mail address}: j.berdermann@desy.de}
\\ \\
\normalsize $^1$Instytut Fizyki Teoretycznej, Uniwersytet Wroc{\l}awski, Poland,\\
\normalsize $^2$Bogoliubov Laboratory for Theoretical Physics, JINR Dubna, Russia,\\
\normalsize $^3$DESY Zeuthen, Germany
}
\date{\normalsize December 29, 2011}

\begin{document}

\maketitle
\begin{abstract}
A combined effective model reproducing the equation of state of hadronic matter
as obtained in recent lattice QCD simulations is presented.
\end{abstract}

\section {Introduction}

We are going to construct\footnote{A more elaborate, detailed description of
this work will be presented elsewhere\cite{paper}.} a combined effective model
reproducing the equation of state of hadronic matter as obtained in recent
lattice QCD simulations \cite{Borsanyi:2010cj,Bazavov:2009zn}.
The model should reproduce basic physical characteristics of processes
encountered in the dense hadronic matter, from the hot QCD phase through the
critical temperature region till the lower temperature hadron resonance gas
phase.
It has been shown in \cite{Karsch:2003vd} that the equation of state derived
from that time QCD lattice calculation \cite{Karsch:2001cy} can be reproduced
by a simple hadron gas resonance model below  critical temperature $T_c$.
For higher temperatures the model is modified by introducing finite widths of
heavy hadrons \cite{Blaschke:2003ut} with a heuristic ansatz
for the spectral function which reflects medium modifications of hadrons.
This Mott-Hagedorn type model is constructed to fit nicely the lattice data,
also above $T_c$ where it does so because it leaves light hadrons below a mass
threshold of $m_0=1$ GeV unaffected.
The description of the lattice data at high temperatures is accidental because
the effective number of those degrees of freedom approximately coincides with
that of quarks and gluons.
In order to remove this unphysical aspect of the otherwise appealing model
one has to extend the spectral broadening also to the light hadrons and thus
describe their disappearance due to the Mott effect while simultaneously the
quark and gluon degrees of freedom appear at high temperatures due to chiral
symmetry restoration and deconfinement.
In the present contribution we will report first results obtained by
introducing a unified treatment of all hadronic
resonances with a state-dependent width $\Gamma_i(T)$ in
accordance with the inverse hadronic collision time scale from a recent model
for chemical freeze-out in a resonance gas \cite{Blaschke:2011ry}.
The appearance of quark and gluon degrees of freedom is introduced by the
Polyakov-loop improved Nambu--Jona-Lasinio (PNJL) model
\cite{Fukushima:2003fw,Ratti:2005jh}.
The model is further refined by adding perturbative corrections to
$\mathcal{O}(\alpha_s)$ for the high-momentum region above the three-momentum
cutoff inherent in the PNJL model.
One obtains eventually a good agreement with lattice QCD data, comparable with
all important physical characteristics taken into account.

\section{Extended Mott-Hagedorn resonance gas}

We introduce the width $\Gamma$ of a resonance in the statistical
model through the spectral function
\begin{equation}
\label{one}
A(M,m)=N_m \frac{\Gamma \cdot m}{(M^2-m^2)^2+\Gamma^2 \cdot m^2}~,
\end{equation}
where $N_m$ is the standard normalization factor.

The model ansatz for the resonance width $\Gamma$ is given by
\cite{Blaschke:2003ut}
\begin{equation}
\label{three}
\Gamma (T) = C_{\Gamma}~ \left( \frac{ m}{T_H} \right)^{N_m}
\left( \frac{ T}{T_H} \right)^{N_T} \exp \left( \frac{ m}{T_H }
\right)~,
\end{equation}
where $C_{\Gamma} = 10^{-4}$, $N_m = 2.5$, $N_T = 6.5$ and $T_H = 165$ MeV.

For simplicity, we assume $n_{S} = 0$ for the strangeness number density and
$n_{B}= 0$ for the baryon number density.
Then the respective chemical potentials $\mu_{B} = 0$ and $\mu_{S} = 0$ always,
so that the temperature is the only significant statistical parameter here.

The energy density of this model can then be cast in the form
\begin{equation}
\label{four}
\varepsilon(T) =
\sum_{m_i < m_0} g_i\,\varepsilon_i (T;m_i)
+ \sum_{m_i \geq m_0} g_i\,\int_{m_0^2}^\infty {d(M^2)}
\,A(M,m_i)\,\varepsilon_i (T;M),
\end{equation}
where $m_0 = 1$ GeV and the energy density per degree of freedom with a mass
$M$ is
\begin{equation}
\label{five}
\varepsilon_i (T;M)  = \int  \frac{d^3 k}{ (2 \pi)^3 }
\frac{\sqrt{k^2+M^2}}{\exp \left(\frac{\sqrt{k^2 +M^2}}{T}
\right)  + \delta_i } \, ,
\end{equation}
with the degeneracy  $g_i$.
For mesons, $\delta_{i} = -1$ and for baryons $\delta_{i} = 1$.
For our hadronic gas kept at fixed volume we have the relation
\begin{equation}
\label{six}
\varepsilon = T \cdot s - P = T \cdot \frac{\partial P}{\partial T} - P\,,
\end{equation}
where $P = P(T)$ and $s = s(T)$ are the pressure and entropy density.

\begin{figure}[!th]
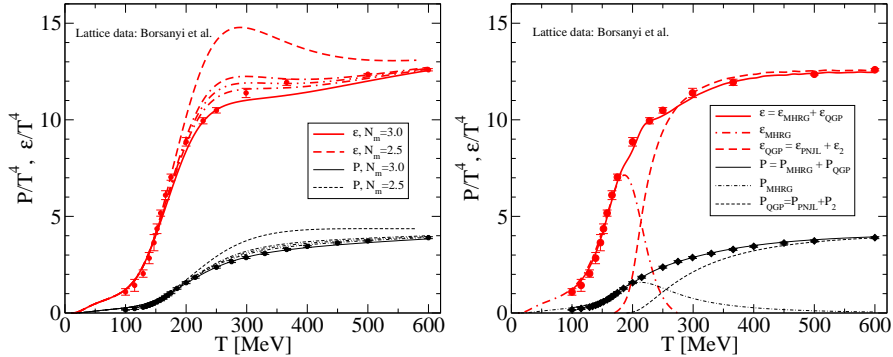

\includegraphics[width=0.48\textwidth]{MHRG_Latt_old}
\includegraphics[width=0.48\textwidth]{MHRG_Latt_PNJL}
\caption{\label{Fig.1}
Left:
Thermodynamic quantities for the old Mott-Hagedorn Resonance Gas model
\cite{Blaschke:2003ut}.
Different line styles correspond to different values for the parameter
$N_m$ in the range from $N_m=2.5$ (dashed line) to $N_m=3.0$ (solid line).
Lattice QCD data are from Ref.~\cite{Borsanyi:2010cj}.
Right:
Thermodynamic quantities for the new Mott-Hagedorn Resonance Gas where
quark-gluon plasma contributions are described within the PNJL model
including $\alpha_s$ corrections (dashed lines).
Hadronic resonances are described within the resonance gas with finite width,
as an implementation of the Mott effect (dash-dotted line).
The sum of both contributions (solid lines) is shown for the energy density
(thick lines) and pressure (thin lines) in
comparison with the lattice data from \cite{Borsanyi:2010cj}.
}
\end{figure}
In the left panel of Fig.~\ref{Fig.1} the results for the pressure and energy
density of the model at this stage are shown.
Although providing us with an excellent fit of the lattice data, the
high-temperature phase of this model is unphysical.
Imposing that all mesons lighter than $m_0=1$ GeV are stable provides us with
a SB limit at high temperatures which mimicks that due to quarks and gluons in
the case for three flavors \cite{Brown:1991dj}.
In reality, due to the chiral phase transition at $T_c$, the quarks loose their
mass and therefore the threshold of the continuum of quark-antiquark scattering
states is lowered.
The light meson masses, however, remain almost unaffected by the increase in
the temperature of the system.
Consequently, they merge the continuum and become unbound - their spectral
function changes from a delta-function (on-shell bound states) to a
Breit-Wigner-type (off-shell, resonant scattering states).
This phenomenon is the hadronic analogue \cite{Zablocki:2010zz} of the
Mott-Anderson transition for electrons in solid state physics
(insulator-conductor transition).
It has been first introduced for the hadronic-to-quark-matter transition in
\cite{Blaschke:1984yj}. Later, within the NJL model, a microscopic approach to
the thermodynamics of the Mott dissociation of mesons in quark matter has been
given in the form of a generalized Beth-Uhlenbeck equation of state
\cite{Hufner:1994ma}, see also \cite{Radzhabov:2010dd}.

As a microscopic treatment of the Mott effect for all resonances is presently
out of reach, we introduce an ansatz for a state-dependent hadron resonance
width $\Gamma_i(T)$ given by the inverse collision time scale recently
suggested within an approach to the chemical freeze-out and chiral condensate
in a resonance gas \cite{Blaschke:2011ry}
\begin{equation}
\label{Gamma}
\Gamma_i (T) = \tau_{\rm coll,i}^{-1}(T)
= \sum_{j}\lambda\,\langle r_i^2\rangle_T \langle r_j^2\rangle_T~n_j(T)~,
\end{equation}
which is based on a binary collision approximation and relaxation time ansatz
using for the in-medium hadron-hadron cross sections the geometrical
Povh-H\"ufner law \cite{Povh:1990ad}.
In Eq.~(\ref{Gamma}) the coefficient $\lambda$ is a free parameter, $n_j(T)$ is
the partial density of the hadron $j$ and the mean squared radii of hadrons
$\langle r_i^2 \rangle_T$ obtain in the medium a temperature dependence which
is governed by the (partial) restoration of chiral symmetry.
For the pion this was quantitatively studied  within the
NJL model \cite{Hippe:1995hu} and it was shown that close to the Mott
transition
%the chiral perturbation theory corrections can be safely neglected and
the pion radius is well approximated by
\begin{equation}
r_\pi^2(T)=\frac{3}{4\pi^2} f_\pi^{-2}(T)
=\frac{3M_\pi^2}{4\pi^2m_q}
|\langle \bar{q} q \rangle_{T}|^{-1}~.
\end{equation}
Here the Gell-Mann--Oakes--Renner relation has been used and the pion mass
shall be assumed chirally protected and thus temperature independent.

For the nucleon,  we shall assume the radius to consist of two
components, a medium independent hard core radius $r_0$ and a pion cloud
contribution
%\begin{equation}
$r_N^2(T)=r_0^2+r_\pi^2(T)~,$
%\end{equation}
where from the vacuum values $r_\pi=0.59$ fm and $r_N=0.74$ fm  one gets
$r_0=0.45$ fm.
A key point of our approach is that the temperature dependent hadronic radii
shall diverge when hadron dissociation (Mott effect) sets in, driven basically
by the restoration of chiral symmetry.
%A further simplifying assumption derives from that statement.
As a consequence, in the vicinity of the chiral restoration temperature all
meson radii shall behave like that of the pion and all baryon radii like that
of the nucleon.
%The chiral condensate used in subsequent calculations has a form
%\[-\langle \bar{q} q \rangle_{T} =
%304.8\left[1-\tanh\left(0.002\,T -1\right) \right]\,.\]

The resulting energy density and pressure behavior is shown in the right panel
of Fig.~\ref{Fig.1}.
This part of the model we call Mott-Hagedorn-Resonance-Gas (MHRG).
When all hadrons are gone at $T\sim 250$ MeV, we are clearly missing degrees
of freedom!

\section{Quarks, gluons and hadron resonances}

We improve the PNJL model over its standard versions
\cite{Fukushima:2003fw,Ratti:2005jh} by adding perturbative corrections in
$\mathcal{O}(\alpha_s)$ for the high-momentum region above the three-momentum
cutoff $\Lambda$.
In the second step, the MHRG part is replaced by its final form, using the
state-dependent spectral function for the description of the Mott dissociation
of all hadron resonances above the chiral transition.
The total pressure obtains the form
\begin{equation}
P(T)=P_{\rm MHRG}(T)+P_{\rm PNJL,MF}(T)+P_2(T) ~.
\end{equation}
where $P_{\rm MHRG}(T)$ stands for the pressure of the MHRG model, accounting
for the dissociation of hadrons in hot dense, matter.

The $\mathcal{O}(\alpha_s)$ corrections can be split in quark and gluon
contributions
\begin{equation}
\label{P2}
P_2(T)=P_2^{{\rm quark}}(T) + P_2^{{\rm gluon}}(T)~,
\end{equation}
where $P_2^{{\rm quark}}$  stands for the quark contribution and
$P_2^{{\rm gluon}}$ contains the ghost and gluon contributions.
The total perturbative QCD correction to $\mathcal{O}(\alpha_s)$ is
\begin{equation}
P_2=-\frac{8}{\pi}\alpha_s T^4(I_{\Lambda}^+
+\frac{3}{\pi^2}((I_{\Lambda}^+)^2+(I_{\Lambda}^-)^2)),
\end{equation}
where
$I^{\pm}_{\Lambda}=\int_{\Lambda/T}^{\infty}\frac{{\rm d}x~x}{{\rm e}^x\pm 1}$.
The corresponding contribution to the energy density is given in standard way
by Eq.~(\ref{six}).

We will now include an effective description of the dissociation of hadrons
due to the Mott effect into the hadron resonance gas model by including the
state dependent hadron resonance width (\ref{Gamma}) into the definition of the
HRG pressure
\begin{equation}
P_{\rm MHRG}(T)=\sum_{i}\delta_id_i\!\int\!\frac{d^3p}{(2\pi)^3}dM\,
A_i(M) T \ln\left(1+\delta_i{\rm e}^{-\sqrt{p^2+M^2}/T} \right)\,.
\end{equation}
From the pressure as a thermodynamic potential all relevant thermodynamical
functions can be obtained.
Combining the $\alpha_s$ corrected meanfield PNJL model for the quark-gluon
subsystem with the MHRG description of the hadronic resonances we obtain the
results shown in the right panel of Fig.~\ref{Fig.1} where the resulting
partial contributions in comparison with lattice QCD data from
Ref.~\cite{Borsanyi:2010cj} are shown.

We see that the lattice QCD thermodynamics is in full accordance with a
hadron resonance gas up to a temperature of $\sim 170$ MeV which corresponds
to the pseudocritical temperature of the chiral phase transition.
The lattice data saturate below the Stefan-Boltzmann limit of an ideal
quark-gluon gas at high temperatures.
The PNJL model, however, attains this limit by construction.
The deviation is to good accuracy described by perturbative corrections to
$\mathcal{O}(\alpha_s)$ which vanish at low temperatures due to an infrared
cutoff procedure.
The transition region $170\le T[{\rm MeV}]\le 250$ is described by the MHRG
model, resulting in a decreasing HRG pressure which vanishes at $T \sim 250$
MeV.

We have presented two stages of an effective model description of QCD
thermodynamics at finite temperatures which properly accounts for the fact
that in the QCD transition region it is dominated by a tower of hadronic
resonances.
To this end we have further developed a generalization of the Hagedorn
resonance gas thermodynamics which includes the finite lifetime of hadronic
resonances in a hot and dense medium by a model ansatz for a temperature- and
mass dependent spectral function.

\section*{Acknowledgments}
D.B. wants to thank K.A. Bugaev and K. Redlich for discussions and collaboration. This work was supported in part by the Polish National Science Center (NCN) under contract No. N N202 0523 40 and by the Russian Foundation for Basic Research grant No. 11-02-01538-a (D.B.).

\end{document}